\documentclass[11pt,a4paper]{article}
\pdfoutput=1
\usepackage{jcappub,topcapt}
\usepackage{sirimacro,microtype}

\usepackage{amsmath}
\usepackage{amsfonts}
\usepackage{amssymb}

\usepackage{graphicx}
\usepackage{epsfig}

\usepackage{mathrsfs}
\usepackage{bm}
\usepackage{microtype}

\usepackage{hyperref}

\begin{document}

\title{On the Abundance of Extreme Voids II: A Survey of Void Mass Functions}

\author{Siri Chongchitnan}
\author{and Matthew Hunt}

\affiliation{E. A. Milne Centre for Astrophysics, University of Hull,  Cottingham Rd., Hull, HU6 7RX, United Kingdom.}

\emailAdd{s.chongchitnan@hull.ac.uk}
\emailAdd{m.d.hunt@2012.hull.ac.uk}

\abstract{
The abundance of cosmic voids can be described by an analogue of halo mass functions for galaxy clusters. In this work, we explore a number of void mass functions: from those based on excursion-set theory to new mass functions obtained by  modifying halo mass functions. We show how different void mass functions vary in their predictions for the largest void expected in an observational volume, and compare those  predictions to observational data. Our extreme-value formalism is shown to be a new practical tool for testing void theories against simulation and observation.

}

\maketitle

\section{Introduction}
Galaxy clusters and superclusters are the most massive gravitationally-bound structures on cosmological scales.
Over the past few decades, the abundances and clustering of the most massive galaxy clusters have provided increasingly stringent cosmological constraints complementary to those from the cosmic microwave background.  Previous studies of structure-formation theory have come to the conclusion that the abundance of massive clusters is well approximated by a \ii{halo mass function}  \cite{ps,sheth,tinker,warren,reed}, particularly on the largest mass scales where clusters evolve almost linearly. The mass function refers to an analytic expression for the number count, $n$, as a function of the cluster mass $M$ (this is often expressed as the differential abundance, $\D n/\D M$).

Cosmic voids, on the other hand,  are the largest cosmological structures  \ii{by volume}, and one might expect that they, too, can be modelled by an analogue of the mass function, say, $\D n/\D R$, where $R$ is the effective void radius. In the rest of this paper, we will use the phrase `void mass function' to refer to this quantity.

Frustratingly, such progress for voids is hampered by many obstacles. At the most basic level, there is a lack of consensus for how exactly a void should be defined. Sheth and Van de Weygaert (SVdW) put forward the `shell-crossing' mechanism \cite{svdw} as the void analogue of the  gravitational collapse that leads to the formation of galaxy clusters. In their theory, the threshold linear underdensity for shell-crossing to occur is $\delta_v\simeq-2.7$ (this is the analogue of the critical overdensity $\delta_c=1.686$ for clusters). However, more recent studies have shown that such shell-crossed voids are excessively underdense, small and rare, and the SVdW theory in its original form does not appear to be a good fit to simulation and observation due to the simplistic assumptions in the theory \cite{chan, sdssvoids, jennings,sahlen}. 

Another problem arises at the observational level. Whilst clusters can be observed (and hence counted) via a fairly well-calibrated mass-luminosity relation, no such relation exists for voids, making number counts more difficult to establish. Instead, watershed-type void-finding algorithms are used to reconstruct void boundary from a galaxy distribution \cite{zobov}. Different void finders do not always agree on the void number count and size distribution \cite{nada2}. Whilst cosmology from void number counts has been previously  explored \cite{pisani}, such results are unfortunately very sensitive to the detail on the particular way that voids are defined and detected. 

In a previous work by one of us \cite{sc1}, we argued that, instead of counting the total number of voids in a given volume, the size of the \ii{largest} void is a more practical  quantity that can be used to constrain void theories using  simulation and observation. Our `extreme-void' formalism was based on the exact extreme-value theory presented in  \cite{harrison1,harrison2} in the context of extreme clusters. In \cite{sc1}, we also derived the extreme-void predictions from the SVdW mass function, and demonstrated the sensitivity of the extreme-void predictions to the variation in the threshold underdensity, $\delta_v$, as well as other cosmological parameters.

In this work, we will extend the results of \cite{sc1} by deriving the extreme-void predictions for a range of void mass functions, some of which are proposed here for the first time. We will demonstrate how these mass functions can be constrained and compared using observational data from the Sloan Digital Sky Survey (SDSS) \cite{sdss11}.

Throughout this paper, we will use the Planck+WMAP cosmological parameters as given in \cite{lahav}.

\section{The distribution of extreme voids}
 
We now summarise key concepts and equations in the exact extreme-value formalism for cosmic voids. See \cite{harrison1,harrison2} for the original expositions in the context of galaxy clusters.

Suppose we have a theory which gives us the (logarithmic) void mass function, d$n/$d$\ln R$. The total number of voids, $N\sub{tot}$, in the redshift bin centred $z$, width $\Delta z$,  is given by 
\ba
N\sub{tot}(z)= f\sub{sky} \int_{R\sub{min}}^{R\sub{max,V}}{\textrm{d}R \over R}\int^{z+\Delta z/2}_{z-\Delta z/2} \textrm{d}z  \diff{V}{z} \diff{n}{\ln R}.\lab{Nto}
\ea
where d$V/$d$z$ is the Hubble volume element, and $f\sub{sky}$ is the observed fraction of the sky (conservatively assumed to be 1 in this work). The radius of observable voids is taken to be between the minimum threshold $R\sub{min}$, and below the maximum $R\sub{max,V}$  determined by the volume of the redshift bin. We take $R\sub{min}=10h^{-1}$Mpc in this work. Whilst the number count is sensitive to the choice of $R\sub{min}$, the extreme-void predictions are not, and we have explicitly verified this.

From Eq. \re{Nto}, we can then construct the probability density function (pdf) for voids with radius in the infinitesimal interval $[R,R+dR]$, in the redshift bin centred $z$, width $\Delta z$, as
\ba
f(R,z)= {f\sub{sky}\over N\sub{tot}}  \int^{z+\Delta z/2}_{z-\Delta z/2} \textrm{d}z  \diff{V}{z} {1\over R}\diff{n}{\ln R}.\lab{pd}
\ea
(essentially by removing the d$R$ integral in \re{Nto} and dividing by itself). It is clear that 
$$\int_{R\sub{min}}^{R\sub{max,V}} f(R,z)\,\D R=1$$
for all $z$, and so our $f(R,z)$ does behave like a pdf as expected.
Furthermore, the cumulative probability distribution (cdf), $F(R,z)$, can be obtained by integrating the pdf:
\ba F(R,z)=\int_{R\sub{min}}^R f(r,z)\, \D r.\ea

Now consider $N$  observations of voids drawn from the distribution with cdf $F(R,z)$ from a bin centred at redshift $z$. [It should be understood that all our pdfs and cdfs are redshift dependent, so for convenience we will just write $F(R)$ to mean $F(R,z)$.] We ask: what is the probability that the \ii{largest} void observed will have radius $R^*$? The required probability, $\Phi$, is simply the product of the cdfs:
\ba \Phi(R^*,N)=F_1(R\leq R^*)\ldots F_N(R\leq R^*)=F^N(R^*),\ea
assuming that void radii are independent, identically distributed variables. As $\Phi$ is another cdf, the pdf of \ii{extreme-size} void can be obtained by differentiation:
\ba \phi(R^*,N)=\diff{}{R^*}F^N(R^*)=Nf(R^*)[F(R^*)]^{N-1}.\lab{eev}\ea
Assuming all voids are observed, we take $N=N\sub{tot}$. It is also useful to note that the peak of the extreme-value pdf (the turning point of $\phi$) is attained at the zero of the function
\ba X(R) = (N-1)f^2 +F \diff{f}{R},\lab{XX}\ea
as can be seen by setting $\text{d}\phi/\text{d}R^*=0$.

In summary, starting with the void mass function, d$n/$d$\ln R$, one can then derive the pdf of extreme voids \re{eev}. In this work, we will follow the convention in large-scale structure literature and parametrize the mass function using the  \ii{multiplicity function}, $f(\nu)$, where  $\nu\equiv |\delta_v|/\sigma$, and $\sigma=\sigma_R$ is the rms dispersion of perturbations smoothed on scale $R$. We express the mass function as
\ba \diff{n}{\ln R}&= f(\nu){3\over {4\pi R_L^3}}  \diff{\ln\sigma^{-1}}{\ln R}\bigg|_{R=R_L}.\lab{param} \ea
The linear-theory void radius, $R_L$, is related to the nonlinear radius, $R$, by the condition 
$R_L=R(\rho_v/\rho_m)^{1/3},$
where $\rho_v$ is the energy density within a void, and $\rho_m$ is the background matter density. Assuming approximate spherical evolution of voids, the ratio of the nonlinear underdensity is well-approximated by the relation
\ba{\rho_v/\rho_m}\approx\bkt{1-{\delta_v\over1.594}}^{-1.594}.\lab{consis}\ea
See \cite{bernardeau} and Appendix B of \cite{jennings} for detail. 


We now discuss a range of multiplicity functions, $f(\nu)$, in equation \re{param}.

\subsection{The SVdW mass function}

The SVdW theory \cite{svdw} is based on the so-called `excursion-set' formalism, which, in the context of voids, postulates that voids are formed where the overdensity field $\delta(\mb{x})$ performs a random walk to a value below a threshold barrier, $\delta_v$. The SVdW multiplicity function is given by
\ba f\sub{SVdW}(\nu)&=2\pi x^2\sum_{j=1}^\infty  j e^{-(j\pi x)^2/2}\sin(j\pi\mc{D}),\\ \mc{D}&= \bkt{1+ \delta_c/|\delta_v|}^{-1}, \qquad x=\mc{D}/\nu. \notag\ea 
In practice, we evaluate the SVdW mass function up to $j=12$ in the sum, with additional terms making negligible  difference in the extreme-void predictions.

\subsection{Modified halo mass functions}

The excursion-set approach in fact goes further back to the pioneering work of Press and Schechter \cite{ps}, who used the random-walk formalism to derive the abundance of massive galaxy clusters. Their work has now spawned a great variety of halo mass functions which have been tested against a wide range observations and simulations. Many of these mass functions are semi-analytic, or calibrated using large-scale simulations, assuming certain so-called `universal' form. See, for example,  \cite{lukic,murray} for reviews of the halo mass functions.

In \cite{kamionkowski}, the authors proposed that void mass functions could be derived from a simple modification to halo mass functions. It was based on the following relation  for the probability  that the value of the local density contrast, $\delta_R(\mb{x})$ (smoothed on scale $R$), is above or below a certain threshold:
\ba P(\delta_R(\mb{x})<\delta_v)= 1-P(\delta_R(\mb{x})>\delta_v).\lab{ohyes}\ea
Differentiating \re{ohyes} with respect to $R$ yields, on the LHS, $\D P(\delta<\delta_v)/\D R$, which is proportional to $\D n\sub{voids}/\D R$,  whilst the RHS is proportional to the halo mass function $\D n\sub{halo}/\D R$ with $\delta_c=\delta_v$. We note the fact that ${\D n/\D R}\propto|{\D P/ \D R}|.$ We will refer to void mass function obtained in this way as being a `modified' halo mass function.

In this work, we will study extreme-void abundances from three  of the most widely used halo mass functions:
\be
\mbox{Press-Schechter \cite{ps}}&& \qquad f\sub{PS}(\nu) = \sqrt{2\over\pi}\nu e^{-\nu^2/2},\\
\mbox{Sheth-Tormen \cite{sheth}}&& \qquad f\sub{ST} = 0.322\sqrt{2a\over\pi}\nu\exp\bkt{-{a\nu^2\over2}}\bkts{1+\bkt{a\nu^2}^{-0.3}},\\ && \qquad a= 0.707,\notag\\
\mbox{Tinker \etal\ff \cite{tinker}}&& \qquad f\sub{Tinker} = 0.368\bkts{1+\bkt{\beta\nu}^{-2\phi}}\nu^{2\eta+1}e^{-\gamma\nu^2/2},\\
&& \qquad \beta = 0.589(1+z)^{0.2}, \phi=-0.729(1+z)^{-0.08},\nn\\
&& \qquad \eta = -0.243(1+z)^{0.27}, \gamma = 0.864(1+z)^{-0.01}.\nn
\ee

\section{Model comparison}

Working with the four void mass functions described thus far, we now show how they can be tested against observation, and compared against one another in terms of goodness-of-fit. 

To demonstrate our method, we choose the void threshold underdensity, $\delta_v$, as the only free parameter. Given a mass function, we set out to find the value of $\delta_v$ that best fits the extreme-void observation derived from SDSS data described below. 

Whilst changing $\delta_v$ in the SVdW theory will result in voids that do not undergo shell-crossing, it has previously been shown that changing $\delta_v$ can produce a surprisingly  good fit to simulation \cite{chan,jennings}. It is arguably an ad-hoc parameter to vary, but given the lack of understanding in the mass function for galaxy voids, we believe it is a constructive step towards reconciling void theory with  observation.

\subsection{The data}

We used the publicly-available void data\footnote{\url{http://www.icg.port.ac.uk/stable/nadathur/voids}} from Nadathur \cite{nadathur3} (early 2016 release) who, using a watershed algorithm, identified almost 9000 voids in the SDSS DR11 data using the LOWZ ($0.15\lesssim z\lesssim 0.42$) and CMASS ($0.44\lesssim z\lesssim 0.68$) surveys. We note that uncertainties in the void effective radius (the `error bars') are not part of the published the catalogue.

\subsection{Maximum-likelihood analysis}

\begin{figure}[t]    \centering
   \includegraphics[width=4in]{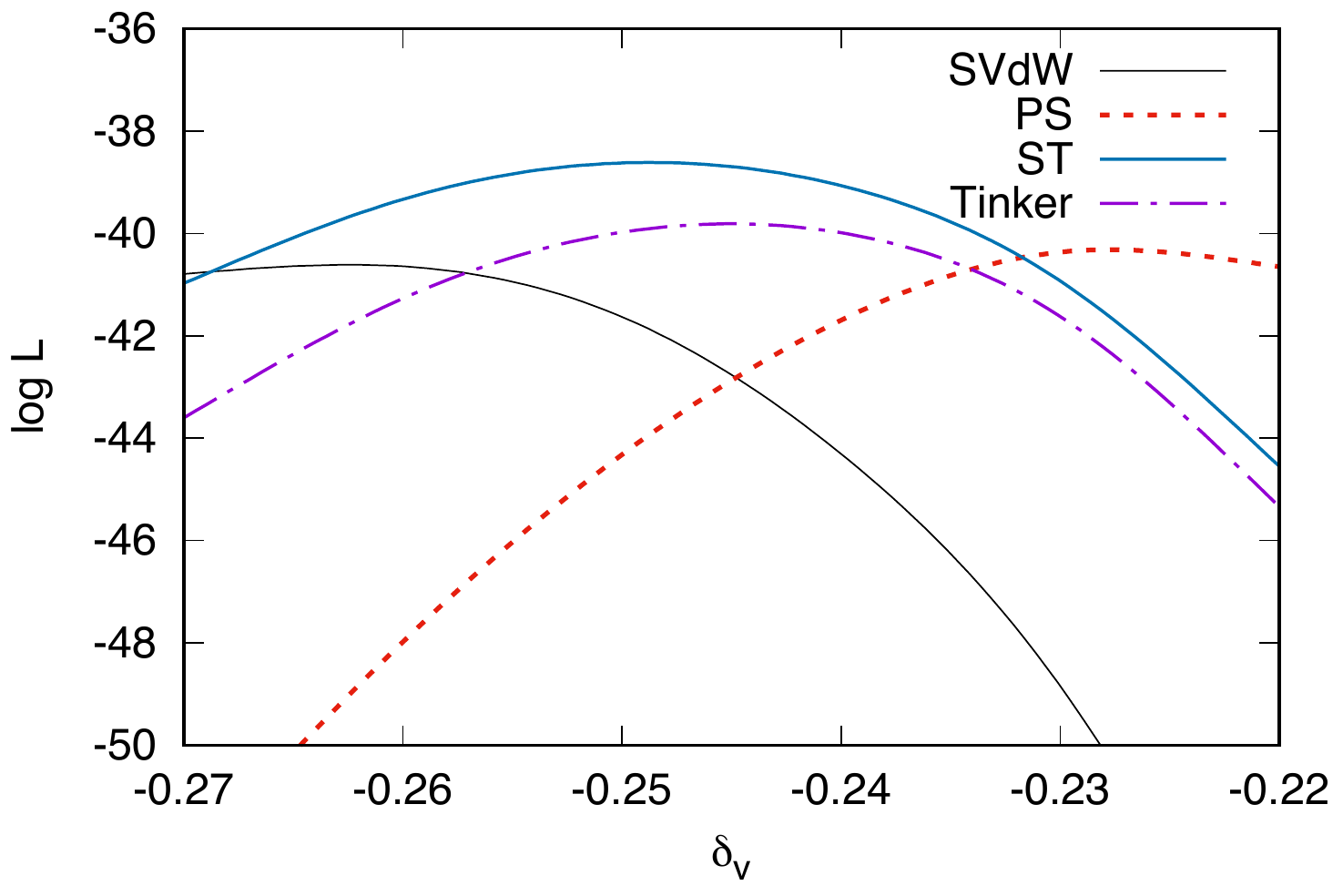} 
   \caption{The log likelihood for various void models as a function of $\delta_v$. The best fit values of $\delta_v$ and maximum values of $\log\mc{L}$ are shown in Table \ref{tabdv}.}
   \label{figlog}
\end{figure}

\begin{table}
   \centering
   \topcaption{Maximum log-likelihood values for various void models.} 
   \begin{tabular}{|c|cccc|}
   \hline
   Model & SVdW & Press-Schechter & Sheth-Tormen & Tinker  \\
   \hline
   Best-fit $\delta_v$ &  $-0.265$ & $-0.23$ & $-0.24$ & $-0.25$\\
   \hline
   $\log\mc{L}$ & $-40.6$ & $-40.4$ & $-39.1$ & $-40.0$\\
   \hline
         \end{tabular}
   \label{tabdv}
\end{table}

We split the void data into 10 bins and identified  the largest void found in each bin. For each of these 10 extreme voids with radius $R^*_i$ ($i=1\ldots10$), we calculate the value of the pdf, $\phi(R^*_i)$, and form the log likelihood function defined as
\ba\log\mc{L}=\sum_{i=1}^{10}\log\phi(R^*_i).\ea
For a given mass function, we determine the value of $\delta_v$ which will maximise the log likelihood. These best-fit values of $\delta_v$ and the corresponding values of $\log \mc{L}$ are shown in Table \ref{tabdv}, and the corresponding graphs of $\log \mc{L}$ as a function of $\delta_v$ are shown in Figure \ref{figlog}.

\begin{figure}    \centering
   \includegraphics[width=6in]{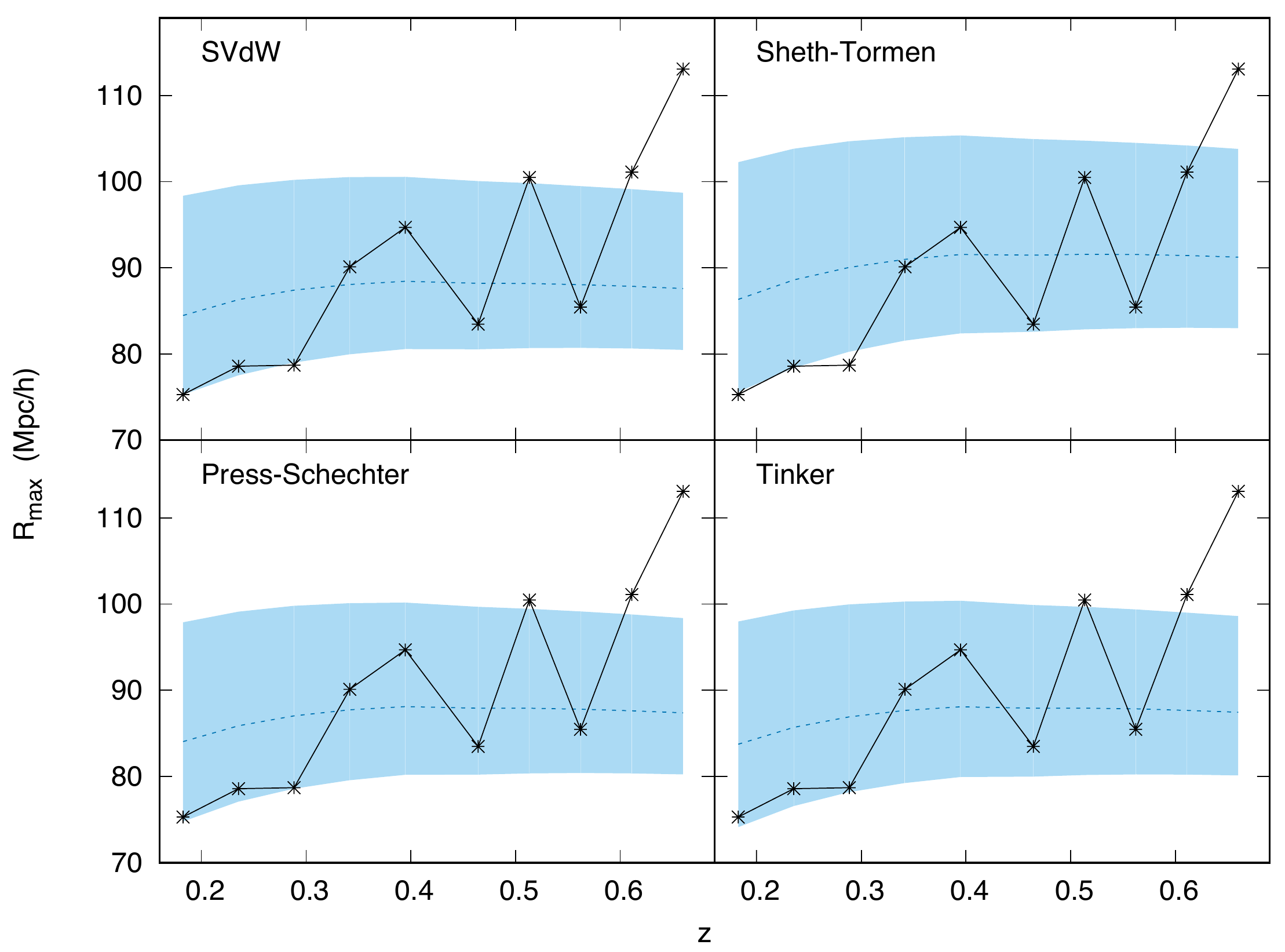} 
   \caption{The extreme void predictions of 4 mass functions using the best-fit values of $\delta_v$ as shown in Table \ref{tabdv}. The data points are from SDSS12 [cite Nadathur]. The band in each figure shows the boundary between the 5th and 95th percentiles of the extreme-void distributions. The median values of the distributions are shown in dotted line near the centre of each band.}
   \label{figpanel}
\end{figure}

In addition, the extreme-void predictions for the four mass functions using these best-fit values of $\delta_v$ are shown in Figure \ref{figpanel}, together with the data points. The shaded region in each of the 4 panels represents the 5th to 95th percentile of the extreme-void distribution (with the median shown with a thin dashed line).  The percentiles are quoted instead of the standard deviation because such pdf has been shown to be strongly non-Gaussian \cite{sc1}. It is also helpful to think of each vertical slice as the top-down view of a pdf, $\phi(R^*)$, at a fixed redshift, peaking near the dashed line. 

In general, each blue band shows a family of curves that peak around $z\sim0.2-0.3$ and thereafter decay very slowly as $z$ increases. This agrees with previous simulations \cite{watson} which found that, at high redshifts, the size of the largest voids is almost redshift-independent.

The results show that all four mass functions can provide a viable fit to the data with $\delta_v\sim-0.2$ to $-0.3$. These values are far from $-2.7$ required in the SVdW theory, but nonetheless we can understand this as follows. 

One of the main predictions of these void mass functions is that the larger the void, the less underdense it is (\iee smaller $|\delta_v|$). This is consistent with the fact that voids identified in the SDSS data are as large as $\sim100$ $h^{-1}$Mpc, and correspondingly, according to the excursion-set approach, they are not highly underdense. We note that some comparably low (and lower) values of $|\delta_v|$ have been reported elsewhere by several groups in search of giant voids \cite{szapudi,inoue}, and so our low value for $|\delta_v|$ is not surprising.

Our analysis also allows us to compare the models with one another. The Sheth-Tormen marginally provides a better fit than the rest of the mass functions: the extreme-value pdfs peak at larger values of $R\sub{max}$ and have a larger spread than the rest (this is apparent in the top-right panel in Figure \ref{figpanel}).

The extreme void in the final bin stands out as a curious outlier. This void lies in the CMASS-South  survey with $z\sim0.64$ and $R=113$ $h^{-1}$Mpc. It appears that this void is located near a high-residual area in the survey mask, and  therefore we should  treat its significance with caution\footnote{We thank Seshadri Nadathur for clarifying the nature of this void for us.}. For the present purpose, it is sufficient to note that its presence (or removal) has negligible effect on the log likelihood. 



\section{The moving-barrier mass function}

As another case study, we now apply our extreme-value analysis to a void mass function which behaves very differently from those in the previous Section. 

The SVdW theory is based on an excursion-set formalism using  fixed barriers. Voids are spherical and typically small (voids of radius $\sim100h^{-1}$Mpc are rare). Smaller voids also tend to be more highly underdense than larger ones. In contrast, voids in dark-matter simulation are identified using a watershed code (\eg\ ZOBOV \cite{zobov}), which typically  combines neighbouring underdense regions  to form large, highly asymmetric voids. The relation between the average underdensity and size of these voids is also the opposite to that of the SVdW theory: larger voids tend to be emptier.

To overcome the above inconsistencies, in \cite{achitouv2, achitouv}, the authors presented an analytical framework based on excursion set formalism with a \ii{moving} barrier, with several major improvements to the original SVdW theory. The analytical distribution of voids obtained was found to be consistent with voids identified by ZOBOV from simulations. The resulting mass function can be expressed as:
\ba f(\nu) &= \sum_{i=0}^{3}T_i, \quad \text{where}\ea
\[
\begin{aligned}[c]
T_0 &= -\nu\sqrt{2a\over\pi}\exp\bkt{-{a\over2}(\nu+\beta)^2}, 
\\
T_2 &=  a\delta_v\beta\bkt{\bar\kappa\,\text{erfc}\bkt{-\nu\sqrt{a\over2}}+T_1},
\end{aligned}
\qquad\qquad
\begin{aligned}[c]
T_1 &= -\bar\kappa\nu\sqrt{2a\over\pi}\bkt{e^{-a\nu^2/2}-{1\over2}\Gamma(0,a\nu^2/2)},\\
T_3 &= -a\beta\bkt{{1\over2}\beta\sigma^2 T_1-\delta_v T_2}.
\end{aligned}
\]
The parameters $a=(1+D_B)^{-1}$ where $D_B$ is the spread of the Gaussian barrier in the theory, $\beta$ is its moving rate; and $\bar\kappa=0.465 a$ for $\Lambda$CDM cosmology. We use the same parameter values as in \cite{achitouv}, namely, $(\beta,D_B)=(-0.1, 0.4)$. We then vary the value of $\delta_v$ and use our formalism to obtain the extreme-void prediction of the theory.

\begin{figure}  

 \centering
   \includegraphics[width=3.5in]{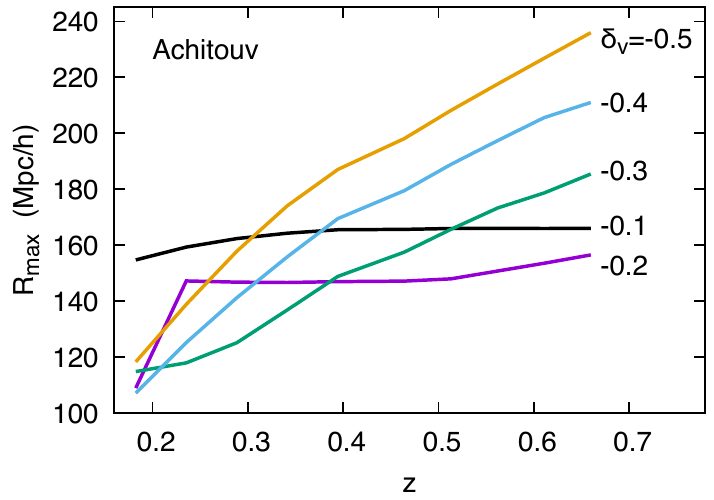} 
   \caption{The peak of the extreme-void pdf $\phi(R)$, as a function of redshift, for the moving-barrier mass function \cite{achitouv2} with $\delta_v$ in the range $-0.1$ to $-0.5$.}
   \label{figachiX}
\end{figure} 

Figure \ref{figachiX} shows the peak of the extreme-value distribution [the solution to equation \re{XX}] as a function of $z$ for a range of values of $\delta_v$. We see that as $|\delta_v|$ increases, the size of the largest voids also (eventually) increases to hundreds of Mpcs, much higher than those typically obtained in the SVdW theory. It is interesting to compare this trend to Figure 3 in \cite{sc1}, which shows the same plot for the SVdW prediction, with curves instead dropping monotonically to tens of Mpcs as $\delta_v$ increases. 

Another interesting behaviour of these $R\sub{max}(z)$ for large $\delta_v$ is the fact that $R\sub{max}$ increases with $z$, meaning that the size of the largest voids at high redshifts is no longer redshift-independent, unlike those for the SVdW and the modified halo mass functions. 

We conclude that the moving-barrier mass function strongly affects the tail of the void size distribution, and the resulting extreme-value distribution is sensitive to the model parameters. We also explored variations in $\beta$ and $D_B$, and found that, like the trend in $\delta_v$, the extreme-value distribution does not change monotonically. Although the fiducial model parameters do not provide a good fit to data, in the future it will be interesting to explore the parameter space $(\delta_v,\beta, D_B)$ further via a Monte-Carlo method.



\section{Summary and discussion}

The primary aim of this work was to explore analytic void mass-functions and compare their predictions for the largest void expected in a given observational volume. In particular, we proposed using void mass functions which are derived from halo mass functions such as those of Press-Schechter, Sheth-Tormen and Tinker \etal.

The exact extreme-value formalism for cosmic voids previously presented in \cite{sc1} was the principal tool used to derive  the pdf of extreme-void radius [Eq. \re{eev}]. We showed that the modified halo mass functions proposed gave similar extreme-void predictions to the SVdW theory.

We demonstrated how our extreme-value framework could be used to compare void mass functions and to fit model parameters.  We showed explicitly how extreme-void data from SDSS could be used to constrain our free parameter, $\delta_v$, using maximum-likelihood analysis. A similar exercise could be used to constrain more parameters via MCMC techniques. This could be applied to, say, fitting the parameters in a `universal'  form of the multiplicity function \cite{tinker,corasaniti}.

In contrast, the moving-barrier mass function was shown to give very different extreme-void predictions from the SVdW and modified halo  mass functions. It typically gives much larger extreme voids, with interesting non-monotonic changes as each model parameter is varied. The voids from this type of model match voids identified in simulation more closely, and it deserves  further investigation.  

Another natural extension of our work is to study the size and redshift of the most massive voids in the Universe given a mass function. A proxy for these quantities is the turning point of the locus of peaks of the extreme-value pdf [as shown in Figure \re{figachiX}], although our framework is adequate for a more complete statistical analysis to be performed, giving a cosmic-void analogue of the most massive clusters in the Universe \cite{holz}.

\bbb

It is both surprising and frustrating that cosmic voids, which are essentially an `inversion' of galaxy clusters, should be so much more difficult to tackle than clusters. The root of the problem appears to be the plurality of voids: voids in dark matter simulations are not the same as galaxy voids identified using biased tracers. The grand challenge is to produce a unifying analytic theory that is flexible enough to predict void distribution for a wide range of void definitions. We believe that the extreme-value formalism demonstrated in this work provides an important and practical first step towards reconciling theory with observation.





\acknowledgments{
We are grateful to Seshadri Nadathur for his insights and for his invaluable work on the SDSS void catalogue. Calculations were performed on the VIPER High-Performance-Computing facilities at the University of Hull.}

\bibliographystyle{jhep}
\bibliography{voidsSCMH}

\end{document}